\documentclass[fleqn,10pt]{wlscirep}
\usepackage[utf8]{inputenc}
\usepackage[T1]{fontenc}
\usepackage{lineno}
\usepackage{lipsum}
\usepackage{graphicx}
\usepackage{float}
\usepackage{caption}
\usepackage{xcolor}
\usepackage{booktabs}

\title{Exploring Gain-Doped-Waveguide-Synapse for Neuromorphic Applications: A Pulsed Pump-Signal Approach}

\author[1,*]{R. Otupiri}
\author[1]{R. Stabile}
\affil[1]{Technische Universiteit Eindhoven, EHCI, Den Dolech 2, 5612 AZ, Eindhoven, Netherlands}

\affil[*]{r.n.otupiri@tue.nl}


\begin{abstract}
Neuromorphic computing promises to transform AI systems by enabling them to perceive, respond to, and adapt swiftly and accurately to dynamic data and user interactions. However, traditional silicon-based and hybrid electronic technologies for artificial neurons constrain neuromorphic processors in terms of flexibility, scalability, and energy efficiency. In this study, we pioneer the use of Doped-Gain-Layer-on-Waveguide-Synapses for bio-inspired neurons, utilizing a pulsed pump-signal mechanism to enhance neuromorphic computation. This approach addresses critical challenges in scalability and energy efficiency inherent in current technologies.

We introduce the concept of Gain on Waveguide Dynamics for synapses, demonstrating how non-linear pulse transformations of input probe signals occur under various pump-probe configurations. Our findings reveal that primarily properties of pulse amplitude, period as well  material properties — such as doping densities and population dynamics — influence strongly the generation of spiking responses that emulate neuronal behaviour and effectively how computational logic is. By harnessing the complex interactions of asynchronous spiking pump techniques and ion densities in excited states, our method produces event-driven responses that mirror natural neuronal functions. This gain-enhanced environment supports short-term memory capabilities alongside essential characteristics like asynchronous spike generation, threshold operation, and temporal integration, foundational to brain-inspired spiking neural network paradigms.
\end{abstract}
\begin{document}

\flushbottom
\maketitle
%
%
\thispagestyle{empty}

\noindent Please note: Abbreviations should be introduced at the first mention in the main text – no abbreviations lists. Suggested structure of main text (not enforced) is provided below.

\section*{Introduction}
The escalating demands of neural networks for computational power and energy have markedly outpaced improvements in electronic hardware, with computational requirements increasing exponentially over the past decade~\cite{1jha2022photonic,2dario2018ai}. While traditional Moore's Law suggests that hardware capabilities double approximately every 18 months, the rapid advancements in AI have driven a steeper increase in computational and energy needs, highlighting the urgent need for more efficient and scalable neuromorphic computing solutions. In contrast, hardware density under Moore's law has exploded approaching a saturation limit while conventional digital computes on the von-Neumann architecture is bottled necked due to the the separation of memory and processor units~\cite{xiang2023photonic}. This in addition to the disparity in the unsustainable consumption of power and carbon emissions has fuelled the pursuit of more energy-efficient neural processing solutions, as traditional computing systems struggle~\cite{2dario2018ai}. The inherent limitations in bandwidth and connectivity further complicate the development of electronic neural hardware~\cite{1jha2022photonic,3nahmias2015excitable}. 

Despite the notable advancements in electronic neuromorphic computing systems utilizing the widely recognized complementary-metal-oxide-semiconductor (CMOS) technology or other emerging electronic devices, these systems continue to encounter significant challenges. Specifically, they are limited by processing speed, energy efficiency, and scalability. Electronic systems often suffer from thermal dissipation issues due to high power consumption, which can hinder their operation and longevity. Additionally, the intrinsic delay in electronic signal propagation and the physical constraints of miniaturization present further barriers to enhancing performance and efficiency~\cite{pei2019towards}.

These challenges have catalysed significant research into neuromorphic photonics, a promising new field that integrates optical technology with neuromorphic designs to create highly efficient, interconnected computing architectures with vast bandwidth capacities.Photonic computing has captured substantial interest due to its intrinsic benefits, including ultra-high speeds, extensive bandwidth, and vast parallelism. These attributes are crucial in addressing the bottlenecks of electronic systems. Photonic devices, by leveraging the speed of light for data transmission and processing, can dramatically reduce latency and energy consumption compared to electronic systems. Furthermore, the ability to multiplex signals across various wavelengths can significantly enhance data throughput and computational density. Various photonic computing methodologies have been explored, driven by their potential to overcome the critical limitations faced by their electronic counterparts~\cite{wetzstein2020inference,shastri2021photonics,huang2022prospects,meng2023compact,jiao2022all,qi2022high}.

Neural network models are typically categorized into two main types: artificial neural networks (ANNs) and spiking neural networks (SNNs)~\cite{4maass1997networks}. ANNs consist of layers of neurons with continuous-valued, non-linear activation functions that process static, analogue inputs~\cite{5wu2018spatio} following the Deep learning. In contrast, SNNs utilize discrete, spatio-temporal spikes as inputs to a computing engine, incorporating the additional time dimension that enhances their computational capabilities and representational capacity compared to ANNs \cite{6maass1996lower,7maass2004computational}. Moreover, the sparse nature of the inputs in SNNs means that neurons activate only during spike events hence tend to have significantly lower power requirements due to reduced neuron activity~\cite{1jha2022photonic}.

Two leading approaches in ANNs are the coherent synaptic networks using Mach-Zehnder Interferometers (MZI)~\cite{khonina2024exploring,tait2017neuromorphic} and the incoherent synaptic networks utilizing microring resonators (MRR)~\cite{guo2021integrated,cheng2017chip,nahmias2020laser}. Both methodologies are extensively studied for their compatibility with CMOS-compatible silicon photonics platforms. Despite their potential, these approaches face significant challenges, including substantial optical losses and difficulties in directly implementing nonlinear computations within the photonic domain~\cite{khonina2024exploring,lee2022photonic,zhang2021optical}.

These limitations underscore the urgent need for alternative strategies that can effectively mitigate the drawbacks of MZI and MRR-based systems. Enhanced designs or novel photonic components may provide the necessary solutions to overcome issues related to optical losses and computational constraints~\cite{khonina2024exploring,lee2022photonic,guo2021integrated,cheng2017chip}. Advancements in these areas are essential for developing more robust and efficient photonic neural networks.

Progress in overcoming these challenges is crucial for the advancement of neuromorphic computing. The ability to perform complex, nonlinear computations using photonic systems is fundamental to the development of next-generation cognitive computing platforms~\cite{xiang2023photonic,bruckerhoff2024probabilistic}. By addressing the current limitations, the field can move closer to realizing highly efficient and scalable photonic neural networks capable of sophisticated cognitive tasks.

Spiking neurons are the essential components of Spiking Neural Networks (SNNs), each interconnected through dynamically adjustable synapses. They are largely categorized into opto-electronic and all-optical systems. Opto-electronic spiking systems primarily utilize semiconductor lasers where optical feedback within laser cavities introduces nonlinearities that enable excitability. Notable implementations include graphene excitable lasers, vertical cavity surface emitting lasers (VCSELs), semiconductor optical amplifiers, and distributed feedback lasers. These systems often employ electrical pumping, with either electrical or optical injection driving the spiking behavior, as modeled by the Yamada framework for lasers with saturable absorbers~\cite{1jha2022photonic,khonina2024exploring,lee2022photonic}. Despite their promise, opto-electronic approaches face challenges such as significant optical losses and difficulties in achieving robust nonlinear computations directly within the photonic domain. In contrast, all-optical excitable devices offer a broader diversity of operational principles, ranging from optically pumped lasers and passive optical cavities to advanced material-enhanced structures like phase change materials. Examples include Q-switched lasers, microring resonators, photonic crystal cavities, and microdisk lasers, which enable excitability through purely optical means without the need for electrical components. This diversity allows for more versatile and potentially lower-loss implementations, facilitating complex nonlinear processing essential for advanced neuromorphic applications. In the architecture of an SNN, spiking neurons execute nonlinear spike activations crucial for temporal data processing and neural information encoding, while synapses perform linear weighting functions that modulate the strength and efficacy of neuronal connections based on synaptic plasticity~\cite{1jha2022photonicbruckerhoff2024probabilistic,zhang2021optical,guo2021integrated,cheng2017chip}.

Most current photonic spiking hardware predominantly employs excitable semiconductor lasers built on III-V platforms\cite{8peng2019temporal,9barbay2011excitability}. These compact laser implementations posses all key characteristics of excitability (all-or-none response, refractory period, delayed response related to STDP) making them closely resemble biological neurons in dynamical behaviour. These systems often experience substantial optical losses due to poor light confinement and high material absorption, making them difficult to scale. Recent advancements have also been made in developing spiking neuron capabilities using silicon-based technologies. These studies integrated phase-change materials on silicon and silicon nitride platforms to create spiking-like behaviours through non-linear pulse transformations\cite{10chakraborty2019photonic,11feldmann2019all}. Despite these innovations, such systems depend heavily on the synchronized coordination of output spike pulses with input data. This synchronization requirement introduces significant challenges, notably the absence of temporal encoding, as well as other challenges such as the abscence of optical gain enhancement, and the lack of asynchronicity, which limits the flexibility in network design~\cite{1jha2022photonic,2dario2018ai,9barbay2011excitability}. Moreover, additional studies have successfully implemented spiking based on excitability through the use of a graphene-embedded silicon microring cavity, characterized by several dynamic features. These features include the asynchronous generation of spikes as a reaction to input disturbances, the capacity for threshold operation which necessitates a specific level of input to initiate activity, and the ability to perform temporal integration and cascadability. These devices take advantage of the nonlinear optical effects inherent in silicon and graphene; we see this in how the third order Kerr effect in graphene amplifies the first order dispersion effect in silicon and  and also provides a strong the first order absorption effect known as saturable absorption which induces a nonlinear dependence of the cavity quality factor on intensity~\cite{ataloglou2018nonlinear}. Nevertheless, a significant hurdle associated with these technologies is the high initial input power required to operate and scale these devices, presenting a substantial challenge in their practical application~\cite{1jha2022photonic,2dario2018ai}.

In this paper, we propose an erbium doped gain-layer on waveguide synapse leveraging the optical pulse transformation through a gain-layer-on-waveguide to achieve spiking. Our  approach aims to leverage the complex interactions of asynchronous spiking pump techniques and ion densities in excited states, creating event-driven responses that mimic natural neuronal functions in a gain-enhanced environment with short term memory capability in addition to key characteritics of asynchronous generation of spikes, threshold operation and temporal integration that underpin any brain-inspired computing paradigm. By incorporating these thin gain-layers into photonic circuits, we enhance light-matter interactions, significantly boosting light modulation efficiency—a key factor for scaling and accurately mimicking the asynchronous spike communications seen in neural networks. The intrinsic gain property of these layers also promote the self-sustainability of spikes, paving the way for dynamic, arbitrary photonic neural networks.

\begin{figure}[ht]
\centering
\includegraphics[scale=0.4]{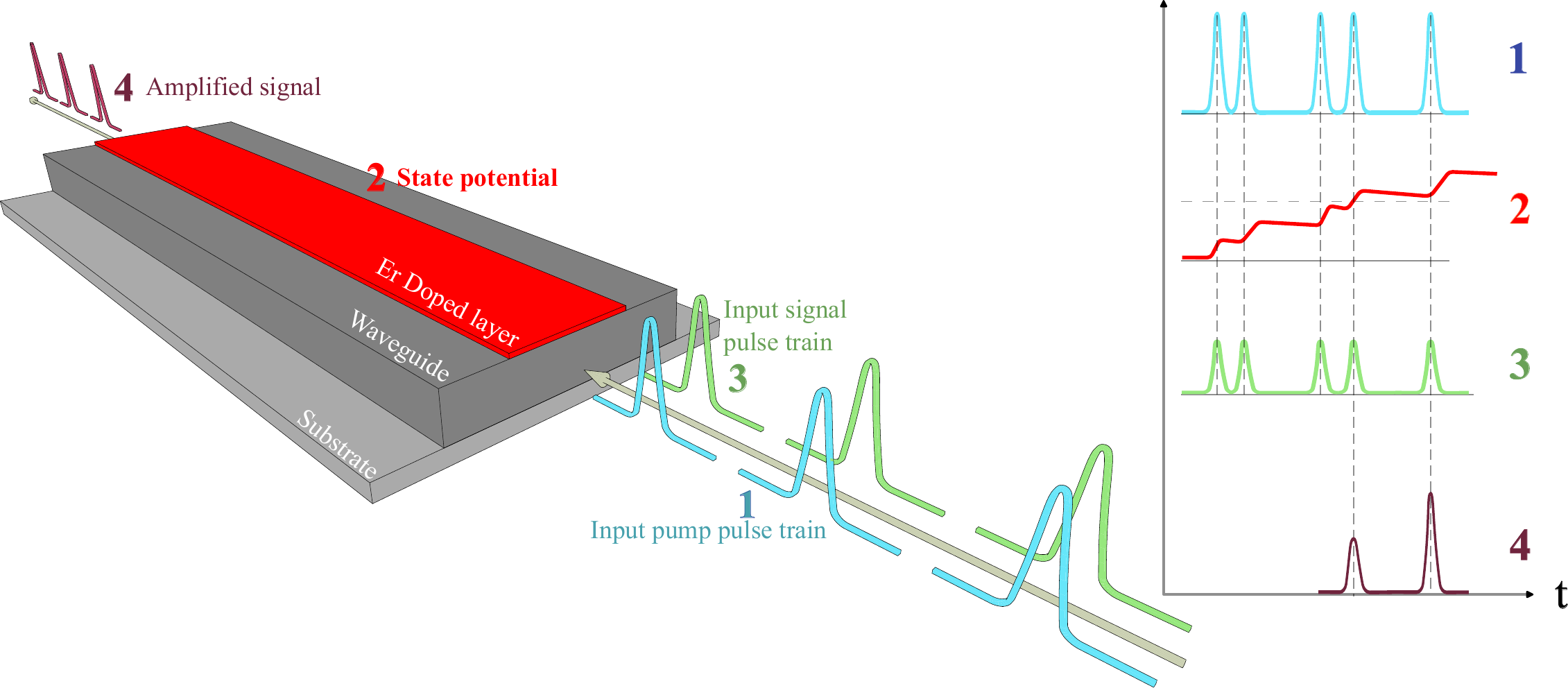}
\caption{A sketch of the spike generation operation and structure of a gain layer on waveguide synapse showing a train of input pump pulse (1-blue curves as pump and 3-green curves as probe)s being injected into the synapse consequentially altering the state potential of the doped layer (2-red curve) to rise above the its transparency threshold resulting in the  registration of a series enhanced signals at the output (4-burgandy curve) from the input signal pulse train (3- green).}
\label{sketch_gain_layer_onwaveguide}
\end{figure}

The Doped-Gain-Layer-on-Waveguide-Synapse represents our groundbreaking advancement in developing spiking neurons by integrating a rare-earth gain layer directly onto a photonic waveguide, as illustrated in Figure~\ref{sketch_gain_layer_onwaveguide}. The prototype features a meticulously layered configuration, beginning with a sturdy substrate base, followed by an efficient photonic waveguide, and topped with the rare-earth-doped gain layer, all within a compact on-chip footprint of approximately one-tenths in height, ones in width, and hundreds in length~\cite{ronn2020erbium}. This sophisticated structure enables the dynamic generation of neuronal spikes through a seamless process: a sequence of input pump pulses (depicted by blue curves) and probe pulses (green curves) are injected into the synapse, which modulates the state potential of the doped layer (orange curve) beyond its transparency threshold. Once this threshold is surpassed, a nonlinear optical response is triggered, resulting in the emission of an output spike (burgundy curve) and a cascade of enhanced signals. This effectively transforms the input pulse train into robust, neuron-like spiking activity, closely mimicking the all-or-none firing behavior of biological neurons. By leveraging the unique properties of the rare-earth gain layer, our Doped-Gain-Layer-on-Waveguide-Synapse not only achieves high-fidelity spike generation but also enhances the synapse’s dynamic response capabilities, paving the way for more efficient and scalable neuromorphic computing systems with minimal spatial requirements suitable for large-scale integration.
The integration of a gain layer marks a substantial advancement in neuromorphic photonics, promising revolutionary changes in how neural networks are designed and implemented.

In this paper we first analyse the behaviour of rare-earth gain-layer-on-waveguide activated by a pulsed pump, delving into the dynamic responses essential for neuromorphic computing. We delve deep into the temporal dynamics of ions across multiple energy levels, exploring both conventional and novel pulsed pump and probe methodologies.

\section*{Temporal Ion Dynamics Modeling}

\begin{figure}[ht]
\centering
\includegraphics[scale=0.15]{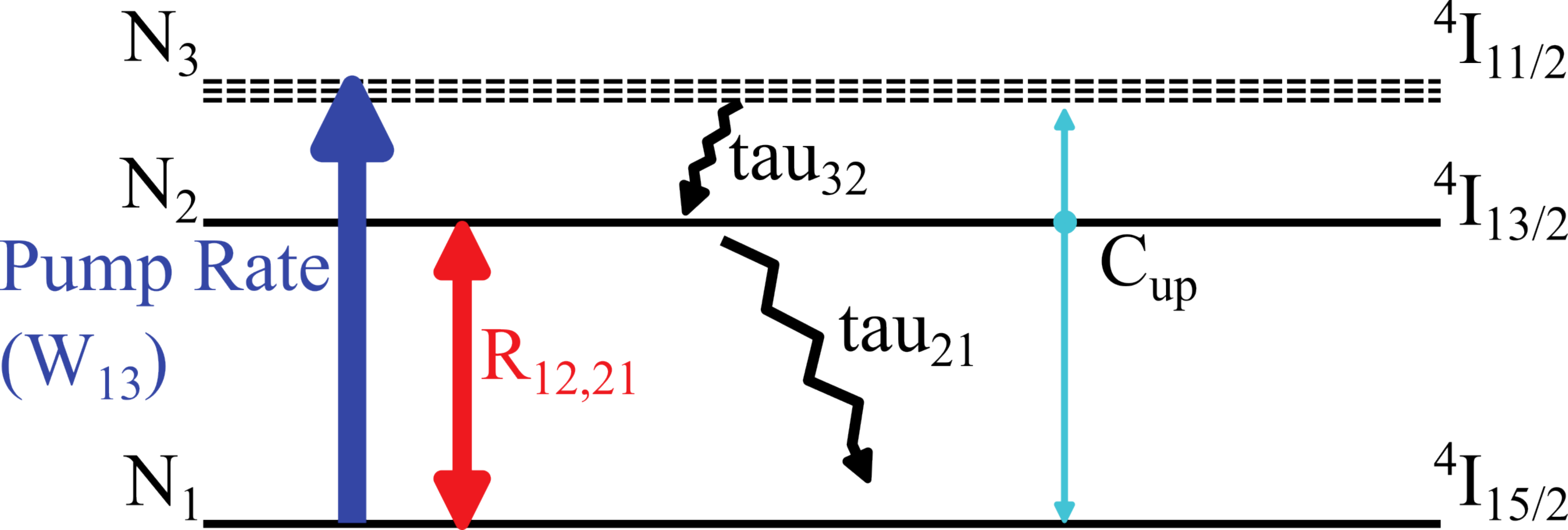}
\caption{Energy-level transitions for the \(\mathrm{Er}^{3+}\) doped three-level system at dual wavelength transition of 980nm and 1550nm.}
\label{Fig:erbium_level_diagram}
\end{figure}

The analysis of temporal dynamics employs an Erbium-doped (\(\mathrm{Er}^{3+}\)) three-level model depicted in Figure~\ref{Fig:erbium_level_diagram}, with a chosen dual pump and probe signal wavelengths of 980nm and 1550 nm, respectively. The model in this study incorporates cross-relaxation effects related to energy transfer within the \(\mathrm{Er}^{3+}\) system. It can also be presented to consider the cross-relaxation with a second ion such as ytterbium for the purposes of co-doping if necessary.
The absorbed pump photon excites ions from a lower energy level to a higher one, where energy is then transferred to the  \({}^{4} I_{11 / 2}\) level of Er3+. There is also the possibility of energy transfer in the opposite direction. The ions in the  \({}^{4} I_{11 / 2}\) energy level of Er3+ can relax to lower levels such as  \({}^{4} I_{13 / 2}\) or  \({}^{4} I_{15 / 2}\), accompanied by photon emission between these levels. 
The population densities of the  \({}^{4} I_{11 / 2}\) and  \({}^{4} I_{15 / 2}\) levels in Er3+ also include contributions from up-conversion at the  \({}^{4} I_{11 / 2}\) level. Although the \(\mathrm{Er}^{3+}\) system is technically a four-level system, the contribution from the fourth level is often negligible due to the rapid decay from the  \({}^{4} I_{15 / 2}\) state to the  \({}^{4} I_{11 / 2}\) state, making the effect almost instantaneous. To simplify, the levels  \({}^{4} I_{15 / 2}\),  \({}^{4} I_{13 / 2}\), and  \({}^{4} I_{11 / 2}\) of Er3+ are labelled as levels 1, 2,and 3, respectively, and their population densities are denoted as \(N_{1}, N_{2}\), and \(N_{3}\). The fundamental equations used used for our model are based on a similar approach in considering a three level system as presented in Karasek~\cite{karasek1997optimum} and Berkdemir~\cite{berkdemir2008numerical}. The rate equations governing these population densities are then formulated based on this system. These are given as:
\begin{equation}
\centering
\begin{aligned}
\frac{\mathrm{d} N_{1}}{\mathrm{d} t} &= -W_{13} N_{1} + \frac{N_{2}}{\tau_{2}} + C_{\text{up}} N_{2}^{2} + W_{31} N_{3} + R_{21} N_{2} - R_{12} N_{1}, \\
\frac{\mathrm{d} N_{2}}{\mathrm{d} t} &= -\frac{N_{2}}{\tau_{2}} + \frac{N_{3}}{\tau_{3}} - 2 C_{\text{up}} N_{2}^{2} + R_{12} N_{1} - R_{21} N_{2}, \\
\frac{\mathrm{d} N_{3}}{\mathrm{d} t} &= W_{13} N_{1} - \frac{N_{3}}{\tau_{3}} - W_{31} N_{3}, \\
\frac{\mathrm{d} S_{}}{\mathrm{d} t} &= (R_{21} N_{2} - R_{12} N_{1})-\text{loss coefficient} \times S  + W_{13} N_{1} - \frac{N_{2}}{\tau_{2}}
\end{aligned}
\label{eq:rateequationsforerbium} 
\end{equation}
Here, \(N_{1}, N_{2}\), and \(N_{3}\) denote the concentrations of \(\mathrm{Er}^{3+}\) ions in the energy levels \({}^{4} I_{15 / 2}\), \({}^{4} I_{13 / 2}\), and \({}^{4} I_{11 / 2}\) respectively, with \(N_{0}\) indicating the total Erbium concentration. \(\tau_{2}\) and \(\tau_{3}\) are the luminescence lifetimes of the respective energy levels, and \(C_{\text{up}}\) represents the CUC coefficient. The parameters \(R_{pa}, R_{pe}, W_{sa}\), and \(W_{se}\) signify the induced transition rates for the signal and pump as per Berkdemir~\cite{berkdemir2008numerical}.

Using this model, we present the simulation outcomes that illustrate critical features of a spiking neuron within a gain-layer-on-waveguide synapse system. Key characteristics demonstrated include the \textit{asynchronous generation of spikes} triggered by input perturbations and the system's \textit{threshold operation}, which dictates the minimum stimulus required for spike initiation. The model also highlights \textit{temporal integration}, where the synapse accumulates input signals over time to trigger a response and also a \textit{short term memory} window immediately after inversion within which spikes occurring there are optically enhanced. Additionally, we observe \textit{signal amplification}, where the neuron enhances the input signal strength, allowing for clearer transmission across the synapse. These characteristics are discussed in more detail in the next section. The asynchronous operation of the system ensures that spike generation does not require synchronization with a global clock, enhancing its utility in realistic computational environments. Finally, the system's cascadability is showcased, indicating its capability to function within larger neural network architectures without loss of signal integrity or functionality.
\begin{figure}[ht]
\centering
\includegraphics[scale=0.45]{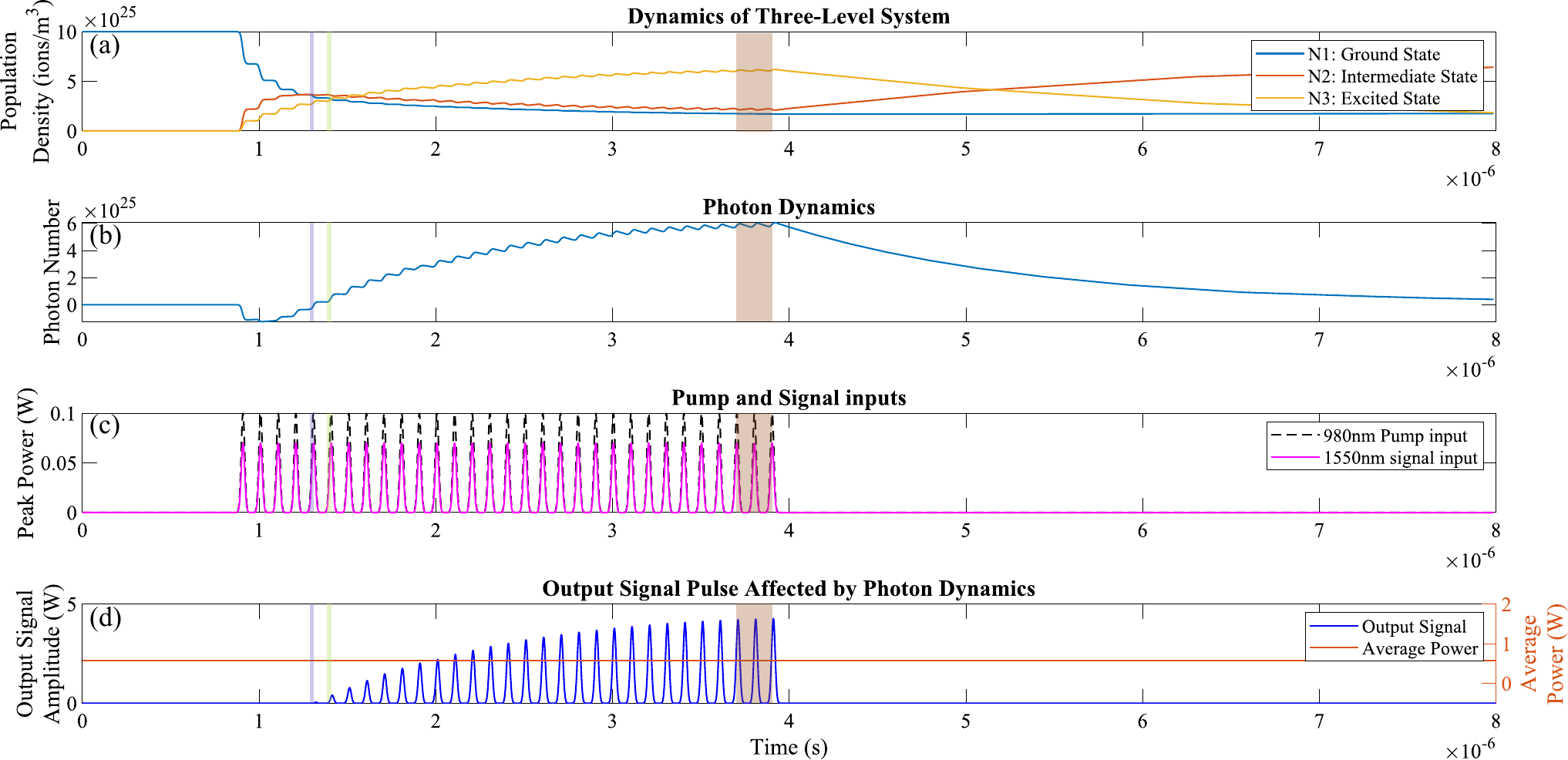}
\caption{This figure illustrates the interplay of population densities across three energy levels, photon generation, and the modulation of input and output signal pulses within a photonic system. Fig~\ref{fig:transparency_saturationthresh_inversion_gain_1}(a) shows the evolving ion concentrations in different energy states, Fig~\ref{fig:transparency_saturationthresh_inversion_gain_1}(b) the dynamics of photon numbers reflecting emission processes, Fig~\ref{fig:transparency_saturationthresh_inversion_gain_1}(c) the temporal profiles of pump and signal pulses, and Fig~\ref{fig:transparency_saturationthresh_inversion_gain_1}(d) the resultant output signal strength and average power, showcasing the system's response to Gaussian pulse inputs.}
\label{fig:transparency_saturationthresh_inversion_gain_1}
\end{figure}
\subsection*{Numerical simulation results}
\begin{figure}[ht!]
\centering
\includegraphics[scale=0.6]{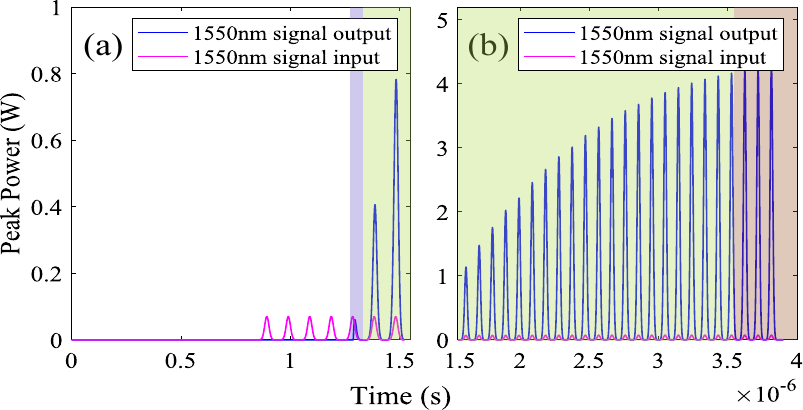}
\caption{Signal Enhancement/Amplification Property in Gain-Integrated Waveguides: The green-shaded region highlights the pronounced signal enhancement achieved through population inversion and gain-layer integration, emphasizing improved output power, signal-to-noise ratio, and pulse fidelity. In contrast, the red-shaded region captures the onset of saturation, revealing the transition from linear amplification to saturation.}
\label{fig:overlay_input_and_output_sig_1}
\end{figure}

To understand the behaviour of the gain-layer-on-waveguide, we inject a burst of optical pulses into the waveguide. As depicted in Figure~\ref{fig:transparency_saturationthresh_inversion_gain_1}, we introduce regular Gaussian pump and probe pulses [Figure~\ref{fig:transparency_saturationthresh_inversion_gain_1}(c)] with peak amplitudes of 0.1W and 0.08W respectively, each lasting 20ns and occurring at intervals of 100ns. Figure~\ref{fig:transparency_saturationthresh_inversion_gain_1}(a) illustrates the resulting dynamics within the three-level ion system, with color-coded regions — blue for transparency, green for inversion, and peach for gain saturation — corresponding to the various threshold states.We can see this better in Figure~\ref{fig:overlay_input_and_output_sig_1}.
Figure~\ref{fig:transparency_saturationthresh_inversion_gain_1}(b) tracks the evolution of the net photon count, which includes photons from both stimulated and spontaneous emissions as the ions transition between levels. This visualization allows us to pinpoint the precise moments of gain saturation, achieve transparency, and reach inversion. 

Notably, following inversion, there is a noticeable increase in optical output peak power also referred to as amplification.The noticeable enhancement of the input pulsed signals, is better illustrated in Figure~\ref{fig:transparency_saturationthresh_inversion_gain_1}(d); here we initially see the absence of any output pulsed signals prior to inversion, a single  output pulsed signal at transparency and enhanced signal pulses post inversion.

To rigorously evaluate the potential of our newly proposed gain-layer-on-waveguide configuration as a foundational element of brain-inspired spiking neural networks, we now present the key characteristics that substantiate its suitability:
\begin{table}[ht]
\centering
\begin{tabular}{@{}ll@{}}
\toprule
Description                                                     	& Parameter and value              \\ 
\midrule
Signal emission cross-section ($m^2$)                              	& $\sigma_{se}$ = 6.8$\times10^{-22}$   \\
Signal absorption cross-section ($m^2$)                            	& $\sigma_{sa}$ = 5.8$\times10^{-22}$   \\
Pump emission cross-section ($m^2$)                                	& $\sigma_{pe}$ = 0                     \\
Pump absorption cross-section ($m^2$)                              	& $\sigma_{pa}$ = 2.58$\times10^{-22}$  \\
Cooperative up conversion coefficient ($m^3$/s)                    	& 4.1$\times10^{-23}$              \\
4I13/2 state lifetime (s)                                       	& $\tau_{32}$ = 3.6$\times10^{-6}$      \\
4I11/2 state lifetime (s)                                       	& $\tau_{21}$ = 7.8$\times10^{-3}$      \\
Waveguide loss/ Photon loss coefficient (s\textasciicircum{}-1) 	& 07$\times10^{5}$                 \\
Channel waveguide core size ($m^2$)              					& 25$\times10^{-12}$               \\ 
Pulse duration ($s$)              									& 20$\times10^{-9}$                \\ 
Pulse interval ($s$)              									& 100$\times10^{-9}$               \\ 
Peak pump power ($W$)   				           					& 10$\times10^{-2}$                \\ 
Signal pump power ($W$)              								& 07$\times10^{-2}$                \\ 
Total concentration of ions ($ions/m^3$)              				& 1$\times10^{26}$                 \\ 
\bottomrule
\end{tabular}
\caption{Simulation parameters used in this work}
\label{tab:parameters}
\end{table}
\noindent
\\ \textbf{Threshold property:} The threshold property is one of the major attributes that should characterise a spiking neuron; this is typically observed as an-all-non response to an expernal perturbation or stimulus input. This means that the system either responds in the form of an excursion/output when the stimulus/input is sufficiently strong or remains non responsive and in its initial state potential when the stimulus/input is not sufficiently strong.
\begin{figure}[ht!]
\centering
\includegraphics[scale=0.4]{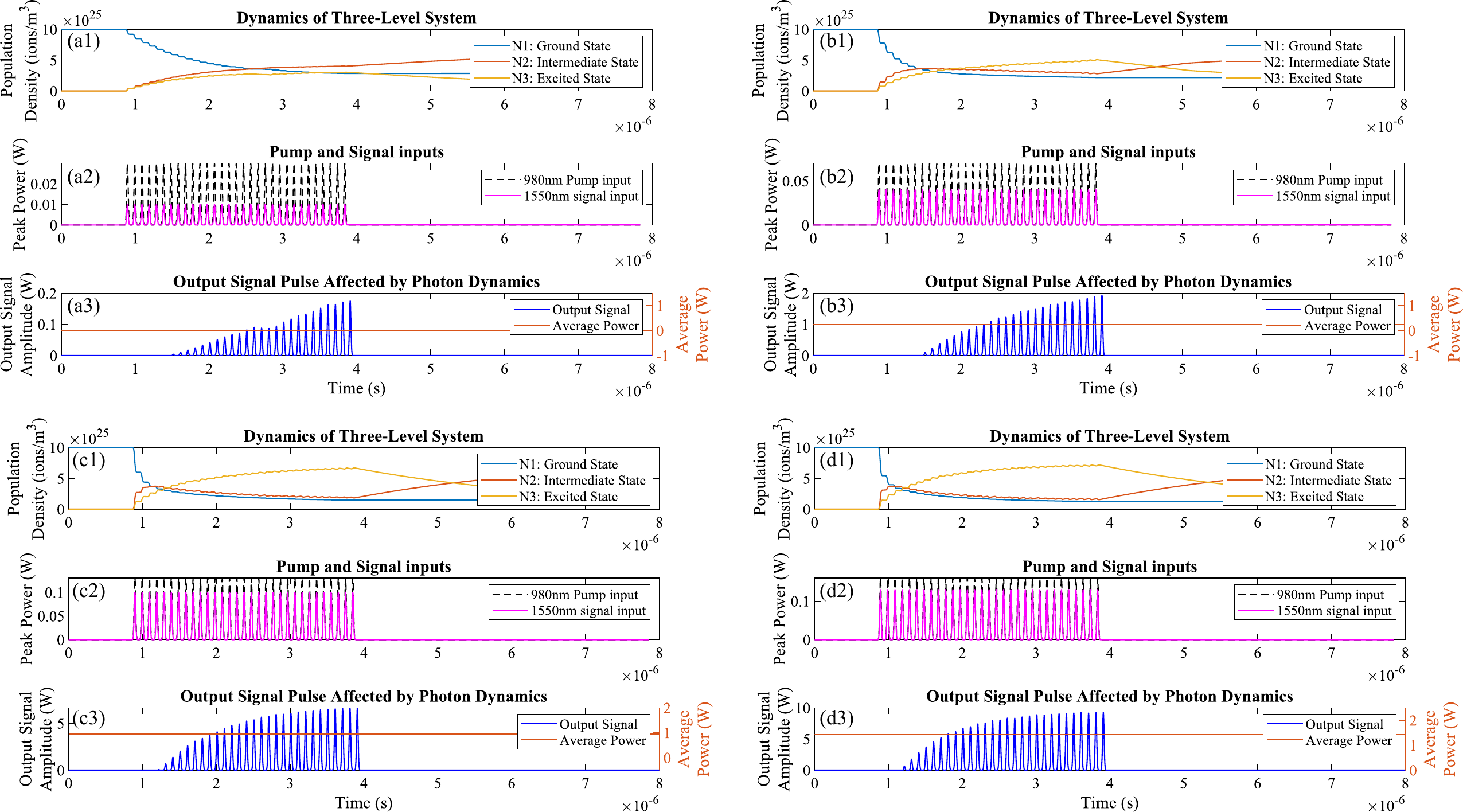}
\caption{ Influence of varying pump and probe input peak powers on the temporal onset of the spiking threshold. Starting from baseline values (pump: 100 mW, probe: 70 mW), the pump and probe peak powers were reduced (30 mW and 10 mW respectively; see (a)) and increased (70 mW and 40 mW respectively; see (b)), while the probe peak power was raised to 130 mW and 100 mW respectively (both shown in (c))and further to 160 mW and 130 mW respectively (both shown in (d)). Increasing peak powers relative to their baselines advances the threshold onset, whereas reducing them delays it, eventually pushing the threshold below the detection limit.}
\label{fig:threshold_property__varyPeakpower}
\end{figure}
\begin{figure}[ht!]
\centering
\includegraphics[scale=0.4]{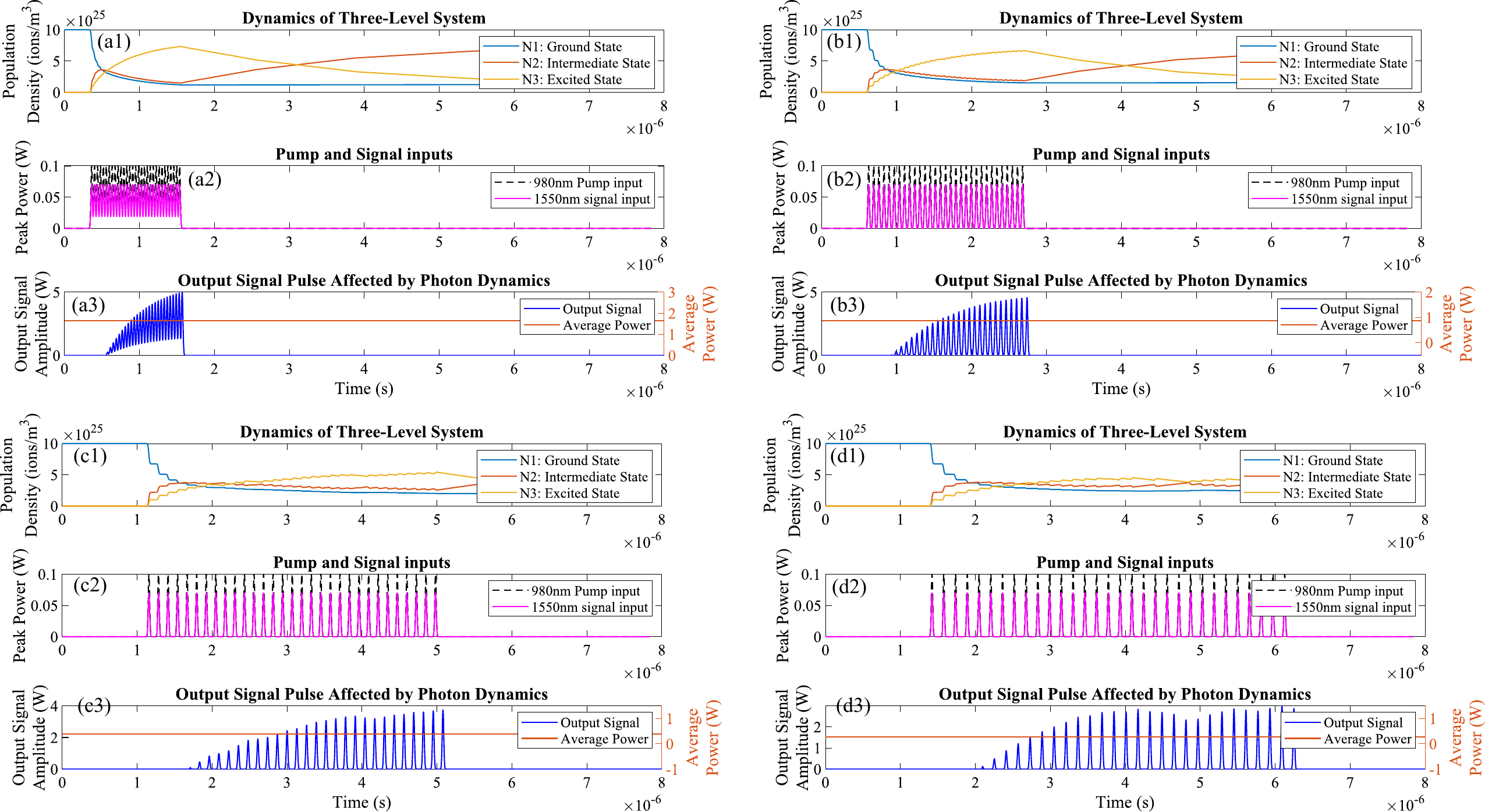}
\caption{Influence of varying the input spike pitch (inter-spike interval) on the temporal onset of the spiking threshold. Each panel corresponds to a different pitch setting: (a) 40 ns, (b) 70 ns, (c) 130 ns, and (d) 160 ns. At shorter intervals (e.g., 40 ns), the device rapidly reaches threshold at the onset of stimulation, yielding poorly defined output spike trains. Increasing the pitch pushes the threshold further away from the initial input, resulting in more clearly defined output spike trains and indicating an optimal operational regime for stable, controllable spiking behaviour.}
\label{fig:threshold_property__varypitch}
\end{figure}
The behaviour of the gain-layer-on-waveguide significantly changes upon reaching the thresholds of transparency and less relevant to this threshold property, inversion. Initially, prior to transparency, there is an absence of any output pulsed signals, indicating that the input signal intensity is below the necessary threshold. This is shown in Figure~\ref{fig:transparency_saturationthresh_inversion_gain_1}(d); the input reaches the transparency threshold, a single output pulsed signal emerges. This transition marks a critical point in the operation of the waveguide, where it begins to effectively process incoming signals.

Figure~\ref{fig:threshold_property__varyPeakpower} illustrates in detail how modifying the input peak power levels of both the pump and probe signals directly influences the temporal onset of the spiking threshold within the system. To establish a baseline, initial pump and probe peak powers of 100 mW and 70 mW, respectively, were selected as reference conditions. From these starting values, the pump peak power was systematically reduced to 70mW and finally to 30 mW (see Fig.\ref{fig:threshold_property__varyPeakpower}(b) and (a) respectively), while the probe peak power was elevated to 130 mW and, subsequently, to 160 mW (see Fig.~\ref{fig:threshold_property__varyPeakpower}(b)).

Notably, the observed threshold onset times displayed a clear dependency on whether the pump and probe powers exceeded or fell below their baseline values. When the input powers surpassed the baseline thresholds, the temporal onset of the spiking event advanced, occurring earlier than under the baseline conditions. Conversely, reducing the pump and probe powers below their respective baseline values delayed the threshold onset. More pronounced reductions in the input powers continued to shift the onset later in time, ultimately driving the threshold below the practical detection limit of the measurement in real terms.

These findings underscore the critical influence of input peak power parameters on the dynamics of the threshold behavior. In particular, even moderate deviations from established baseline powers can substantially alter the timing of key nonlinear phenomena, such as the initiation of spiking. Understanding and controlling these power-dependent effects is therefore essential for designing on the gain-layer-on waveguide which relies on precise threshold timing. By carefully tuning the pump and probe peak powers, one can engineer desired temporal responses, enabling greater control over system performance and stability.

Figure~\ref{fig:threshold_property__varypitch} illustrates in detail how varying the input spike pitch of both the pump and probe signals directly affects the temporal onset of the spiking threshold. As a reference, both the pump and probe pitches were initially set to 100 ns. From this baseline condition, the inter-spike intervals were first reduced from 100 ns to 70 ns and then to 40 ns (see Fig.~\ref{fig:threshold_property__varypitch}(b) and (a), respectively). At shorter interspike intervals (below approximately 70 ns), the device becomes overstimulated almost immediately, causing the threshold to be reached at the onset of the input and resulting in poorly defined output pulse trains (See Fig~\ref{fig:threshold_property__varypitch}(a3)). On the hand, when one approaches wider interspike regimes of 160 ns, the timing of the threshold gets delayed futher from the onset of the inpult sequence of spikes, however, pulsing instabilities start to develop in the output pulse sequence (See Fig~\ref{fig:threshold_property__varypitch}(d3)).

These findings underscore the critical influence of input pulse parameters on the dynamics of the threshold behaviour. In particular, even moderate deviations from established baseline powers can substantially alter the timing of key non-linear phenomena, such as the initiation of spiking. Understanding and controlling these parameters is therefore essential for designing on the gain-layer-on waveguide which relies on precise threshold timing. By carefully tuning the pump and probe peak powers and inter-spike duration, one can engineer desired temporal responses, enabling greater control over system performance and stability.

\noindent
\\ \textbf{Temporal integration property:}
Temporal integration, another fundamental principle required for bio-inspired operation, involves the cumulative summation of multiple, temporally distributed synaptic inputs over a finite interval. This integrative process enables neurons to adaptively modulate their responsiveness and sensitivity to fluctuations in the timing and amplitude of incoming signals. By effectively pooling signals arriving at different moments, temporal integration allows for the stabilization of neural responses, particularly when dealing with noisy or asynchronous inputs. In doing so, it transforms a set of discrete, time-dependent inputs into a more continuous and coherent output.
\begin{figure}[ht!]
\centering
\includegraphics[scale=0.4]{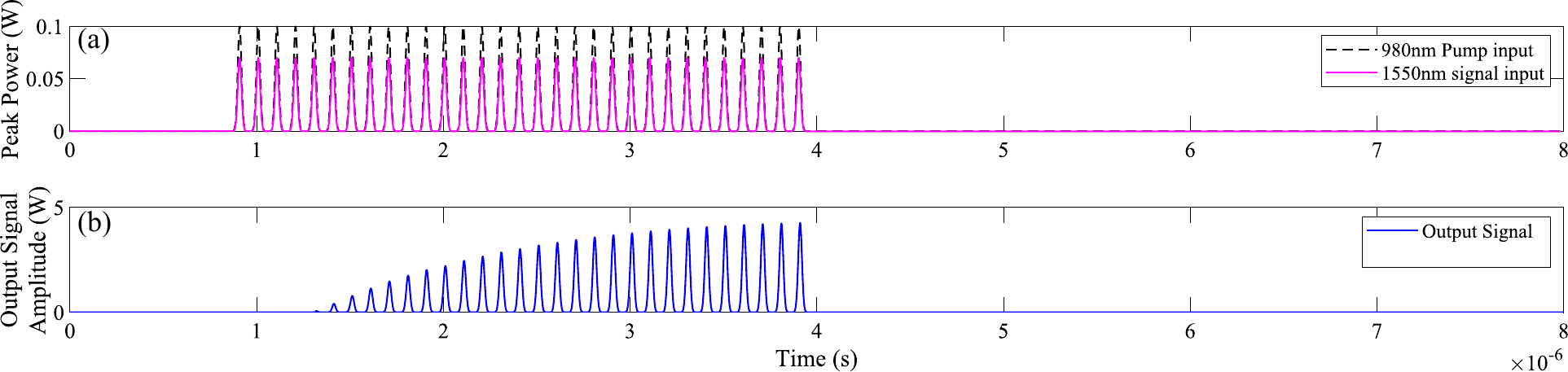}
\caption{Temporal integration within a gain-layer-on-waveguide system. (a) Schematic illustration: sequential optical pulses arriving at staggered intervals enter a waveguide overlaid by a gain medium. As they propagate, the gain layer amplifies their cumulative photon count, effectively integrating temporally distributed inputs. (b) Empirical output signals: the accumulated inputs yield intensified output pulses, demonstrating how dynamic temporal aggregation enhances the overall signal strength and stability.}
\label{fig:temporal_integration}
\end{figure}
A closely related mechanism can be observed in photonic systems employing a gain-layer-on-waveguide configuration, as illustrated in Figure~\ref{fig:temporal_integration}(a). Here, individual optical pulses injected at different time points accumulate within a waveguide that is augmented by a gain medium. As these staggered inputs propagate through the waveguide, the gain layer amplifies the overall photon population. This continuous increase in photon count arises from the effective aggregation of multiple signals that do not coincide precisely in time but overlap within the operational window of the device.

The result of this integration is vividly demonstrated in Figure~\ref{fig:temporal_integration}(b), where the output of the waveguide exhibits intensified optical pulses. The cumulative effect of temporally spaced inputs amplifies the resultant signal, smoothing out fluctuations and producing a more robust output. In essence, the gain-layer-on-waveguide structure mirrors the temporal integration observed in neural circuits: it captures and enhances diverse inputs over a defined timescale, ultimately leading to a strengthened and more stable overall response.
\noindent
\\ \textbf{Signal Enhancement Property:}
Signal amplification is a distinctive feature of the gain-layer-on-waveguide structure, setting it apart from other brain-inspired synaptic systems. In biological synapses, neurons fire, generating signals that are stronger than the associated excitatory inputs. This phenomenon typically arises from the integration of sub-threshold excitatory pulses, followed by a substantial energy release once the threshold is surpassed. This process is conceptually mirrored in the leaky-integrate-and-fire models found in various optoelectronic and all-optical synaptic systems.

Our system, however, operates based on a fundamentally different mechanism. It leverages optically sensitive ion carriers within the node, which enable optical signal enhancement when the system is driven at its specific optical pump wavelength. 

Upon achieving population inversion, the introduction of the gain layer into the waveguide structure results in a significant enhancement of the transmitted optical signal's peak power. Once the population of excited carriers exceeds that of the ground state, leading to inversion, the embedded gain medium amplifies the propagating light substantially.

As a result, the output pulses exhibit a marked increase in intensity, as demonstrated in the green-shaded region of Figure~\ref{fig:overlay_input_and_output_sig_1}. This region highlights the critical role of the gain layer in augmenting the signal amplitude, improving not only the raw output power but also the signal-to-noise ratio and the fidelity of the transmitted pulses.

\begin{figure}[ht]
\centering
\includegraphics[scale=0.44]{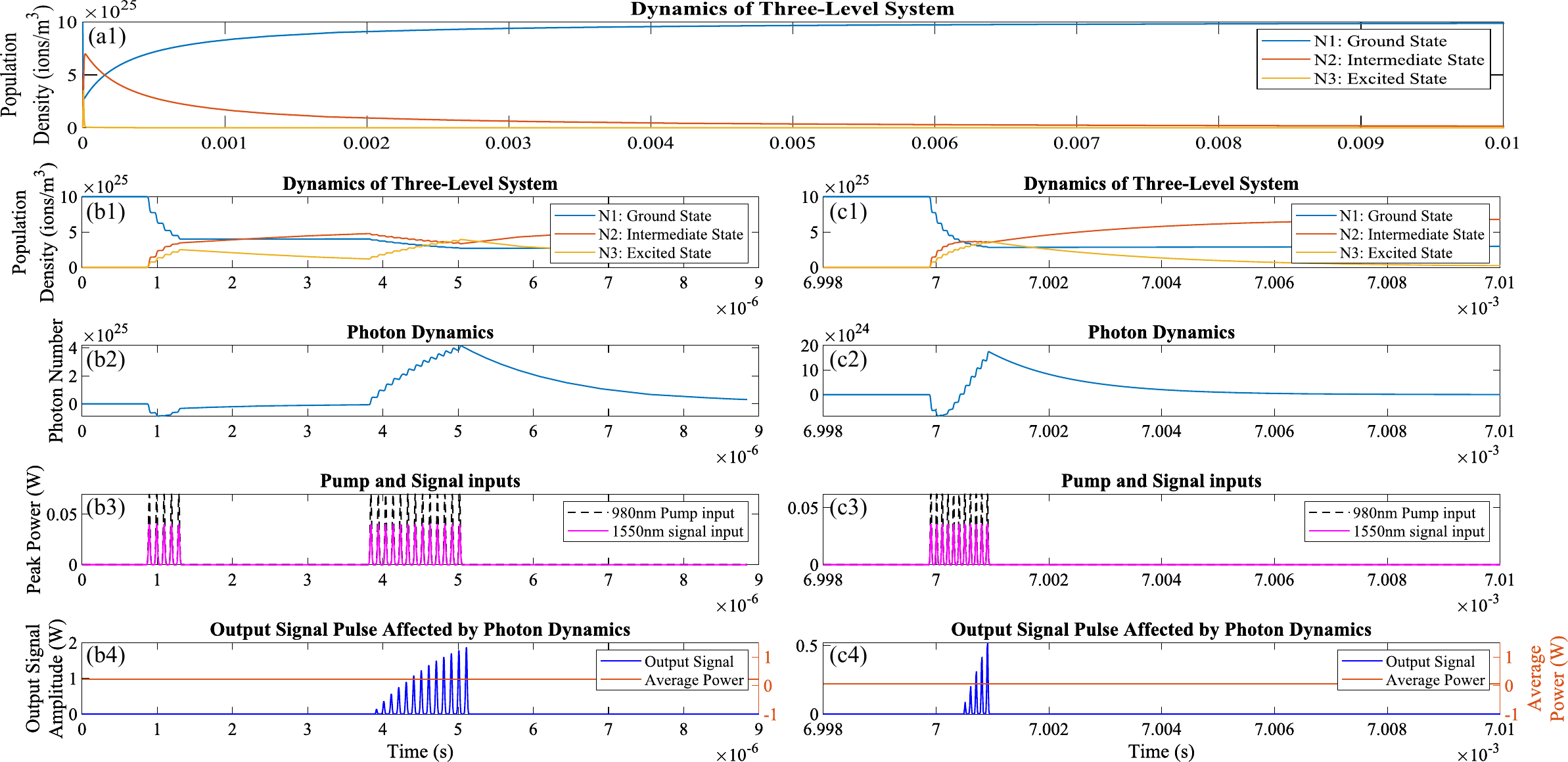}
\caption{This figure illustrates the critical time window of population inversion within the Doped-Gain-Layer-on-Waveguide-Synapse, demonstrating its short-term memory capabilities. Panel (a1) shows the full evolution of ions in the three levels from non-excited to excited and back to non-excited over a 10ms window. Panels (b1) - (b4) demonstrate the enhancement of signal strength following two pulse trains within the decay envelope, showing immediate signal amplification. Panels (c1) - (c4) depict the requirement of additional pulses to achieve inversion and subsequent signal enhancement outside the memory window.}
\label{fig:MERGED_SET1SET2tempmemoryproperty}
\end{figure}

In addition to this direct amplification, the figure also shows the device entering a saturation regime, indicated by the red-shaded region. In this region, the gain medium shifts from linear amplification to a saturation state, where any further increase in input power results in a plateau in output intensity. This saturation effect reveals the inherent limitations of the gain mechanism, where factors such as carrier depletion, spontaneous emission, and thermal loading begin to limit the achievable gain. Collectively, the behaviors observed in the green and red regions of Figure~\ref{fig:overlay_input_and_output_sig_1} demonstrate the operational dynamics of the gain-integrated waveguide device, showcasing its ability to enhance signal clarity and intensity, while highlighting the constraints imposed by material and design limits.

\noindent
\\ \textbf{Signal Enhancement/Amplification Property:}
Furthermore, a pivotal feature of the gain-layer-on-waveguid that is particularly advantageous for neuro-inspired spiking applications is its short-term memory capability. This property allows the waveguide to temporarily maintain its state potential, which can be utilized in the computation and processing of information, similar to how neural synapses function in the brain. Our study, illustrated in Figure~\ref{fig:MERGED_SET1SET2tempmemoryproperty}, delves into this by highlighting the time window during which the waveguide sustains a state of population inversion.	
In Figure~\ref{fig:MERGED_SET1SET2tempmemoryproperty} panels (a1) - (a4), we introduce two sequences of input pulse trains. The initial pulse train surpasses the transparency and inversion thresholds, setting the stage for the subsequent train. This second set of pulses, falling within the decay envelope of the first, immediately experiences signal enhancement. Conversely, in Figure~\ref{fig:MERGED_SET1SET2tempmemoryproperty} panels (b1) - (b4), pulses outside the designated memory window require additional input to first overcome the transparency threshold before any signal enhancement is observed. This disparity in timing for the observation of an amplified spike underscores the waveguide’s capability to differentially respond based on timing, aligning with the Time to First Spike (TTFS) encoding technique we intend to employ in our systems. This approach provides a nuanced insight into how gain-layer-on-waveguide can effectively mimic neural processing dynamics.

This memory window corresponds to the upper state life time of  erbium ions (generally $\sim$10ms). During this period, there is a predominant ion population in the upper energy levels. By understanding and optimizing this inversion duration, we aim to enhance the waveguide's functionality in neuromorphic computing systems, where rapid and transient information processing is crucial.

\noindent
\\ \textbf{Asynchronous operation:}
The final principle which our gain-layer-on-waveguide device posses is asynchronous operation which offers significant advantages in terms of flexibility, efficiency, and real-time processing. In contrast to synchronous systems, which rely on a global clock to coordinate operations, asynchronous systems operate without a unified timing reference, allowing components to function independently. This characteristic is particularly beneficial in mimicking the behaviour of biological systems, where events do not occur at fixed intervals, but are driven by local interactions and stimuli.

\begin{figure}[ht]
\centering
\includegraphics[scale=0.5]{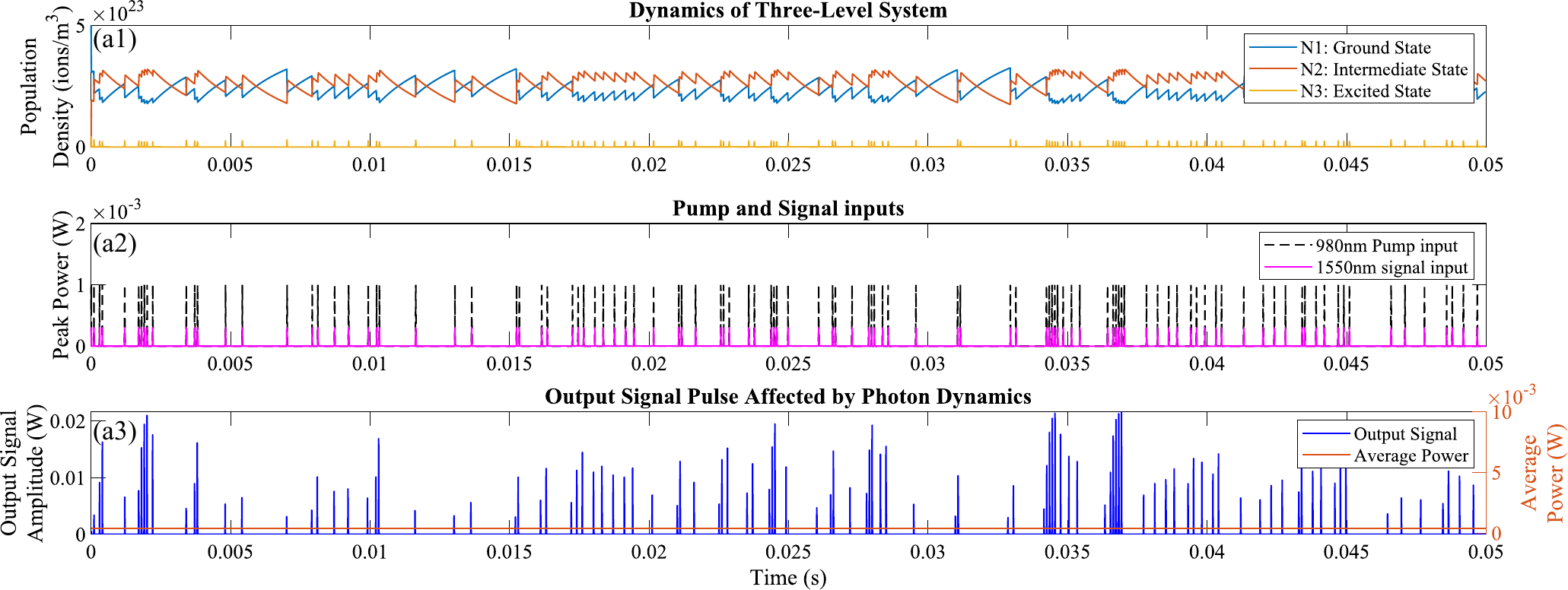}
\caption{Illustration of asynchronous operation in a neuromorphic system, showing independent, event-driven processing where panel (a2) is a temporal sequence from node activates based on local stimuli, without reliance on a global clock. This enables highly parallel, dynamic interactions between components, optimizing efficiency and scalability in signal processing.}
\label{fig:utorandominputsequence}
\end{figure}

Figure~\ref{fig:utorandominputsequence} shows a simulation of independent event driven responses from pre-nodes that have arrived synchronously(See Fig~\ref{fig:utorandominputsequence} (a2)) and triggered  above threshold response operations resembling the dynamic and event-driven nature of biological neurons and synapses(See Fig~\ref{fig:utorandominputsequence} (a3)). Neurons in the brain fire based on local inputs and their internal states, rather than waiting for a global clock signal. This enables highly parallel processing, where each neuron can process information at its own pace, depending on the strength and timing of incoming stimuli. In this context, asynchronous operation minimizes power consumption by ensuring that resources are only engaged when necessary, rather than constantly being refreshed as in synchronous systems.

One of the key advantages of asynchronous operation is its scalability. In traditional synchronous systems, adding more components to the network requires careful management of clock synchronization, which can become increasingly complex as the system grows. In contrast, asynchronous systems scale more naturally because each component operates independently, and the system's overall performance is not limited by the need to synchronize all parts. This property is essential in building large-scale, distributed networks of artificial neurons, where each node operates based on local conditions and the interactions between nodes determine the overall network behaviour.
Asynchronous operation also enhances fault tolerance. Since components in an asynchronous system do not rely on a central clock, the failure of one or more components does not necessarily disrupt the entire system. This decentralized nature makes asynchronous systems inherently more robust, particularly in environments where reliability and continuous operation are critical.

\subsubsection*{Comparing Photonic Spiking Neuron Characteristics}
In this section, we evaluate the gain-layer-on-waveguide synapse alongside various spiking neuron devices, focusing on their application within extensive network architectures. Performance in these networks hinges on several critical factors beyond the basic neuron design, including platform, minimal power consumption, firing rate, and device footprint, as summarized in Table~\ref{fig:Comparing Photonic Spiking Neuron Characteristics}.

We begin by examining the physical size of the devices, a pivotal aspect for edge computing applications. Integrated devices tend to outperform discrete ones in terms of compactness. Passive silicon-based devices, due to silicon's high refractive index, are notably the smallest, with our gain-layer-on-waveguide synapse following closely. This compactness is essential for devices that host all-optical spiking neurons. The firing rate, another vital metric, determines the temporal resolution of spike-based processing. High-speed neuromorphic photonics, capable of GHz range processing speeds, significantly outpace electronic systems. Among the evaluated devices, our gain-layer-on-waveguide synapse shows commendable performance, although it is outperformed by the silicon-based all-optical graphene-based neuron.
Regarding power consumption, there is a marked variance among the devices. Typically, optoelectronic devices, especially laser-based ones, use more power compared to all-optical devices. Among the devices assessed, the gain-layer-on-waveguide synapse emerges as a frontrunner in energy efficiency, capitalizing on gain properties that enhance signal integrity within the network.

While no single spiking device dominates across all metrics, advancements in technologies like the gain-layer-on-waveguide synapse are forging pathways to overcome some of the traditional limitations seen in optoelectronic and all-optical systems. These innovations are poised to improve network cascadability and reduce power losses, boosting the overall effectiveness of neuromorphic photonic networks.
\begin{table}[ht]
\centering
\begin{tabular}{@{}lllll@{}}
\toprule
Device                          & Footprint {[}$\mu m^2${]} & Power {[}mW{]} & Platform     & Firing rate {[}GHz{]} \\ \midrule
DFB Laser               		& \textgreater{}600x200   & 260            & InP          & $\sim{2}$             \\
Graphene laser         		& 40                      & 88             & InGaAsP-C-Si & 4                     \\
Microdisk Laser        		& 25 pi                   & .001           & InAsP-Si     & 1                     \\
RT-PD                  		& 600                     & 3              & InGaAsP      & .002                  \\
VCSEL        					& 40,000                  & 0.1            & Discrete     & 1                     \\
2D PhC                 		& 500                     & $\sim{3}$      & InP          & 0.005                 \\
MRR                    		& 25 pi                   & 4              & Si           & 0.005                 \\
PCM-based cavity       		& 3600 pi                 & 30             & SiN          & 0.02                  \\
Graphene-Si MRR       			& 25 pi                   & 100            & Si           & 40                    \\
Gain-layer-on-waveguide	 	& $\sim$ 375 to 900       & \textgreater{} $\sim{20}$               & SiN          & $\sim{0.01}$              \\ \bottomrule
\end{tabular}
\caption{Comparison of gain-layer-on-waveguide synapse with other spiking neuron devices in large network implementations: Evaluating platform, minimal power consumption, firing rate, and footprint~\cite{1jha2022photonic}}
\label{fig:Comparing Photonic Spiking Neuron Characteristics}
\end{table}

\section*{Exploration of SNN Encoding Schemes}
Neural coding schemes play an essential role in converting sensory input data, such as pixel values, into spike patterns that are relayed to excitatory neurons, facilitating the processing of sensory information~\cite{gerstner2014neuronal,gerstner2002spiking} in neural networks. Research has identified four main types of neural encoding: rate coding, time-to-first-spike (TTFS) coding, phase coding, and burst coding~\cite{auge2021survey,guo2021neural}. Each method operates on unique principles to represent the same input data in different ways, offering varied approaches to how information is processed and transmitted.

For example, rate coding translates the intensity of an input, such as the brightness of a pixel, into a spike count within a specified time period. The greater the intensity, the more frequent the spikes. However, rate coding tends to be inefficient in terms of information transmission speed and processing efficacy~\cite{adrian1926impulses}. This is because it relies on statistical analysis over extended durations and primarily considers the firing rate without accounting for the critical role of precise spike timing, a factor that is significant according to various physiological studies~\cite{gerstner1997neural}.

In contrast, TTFS coding uses the timing of the first spike after a stimulus as a marker of input intensity. Here, stronger inputs result in earlier spikes, providing a potentially quicker and more direct method of conveying information~\cite{johansson2004first}. These methods can speed information transmission and offer energy-efficient single-spike requirements. However, they exhibit poorer robustness against interference, and the sparse firing rates make it challenging to directly extend such networks to multiple layers~\cite{wang2023activeness}.
\begin{figure}[ht]
\centering
\includegraphics[scale=0.035]{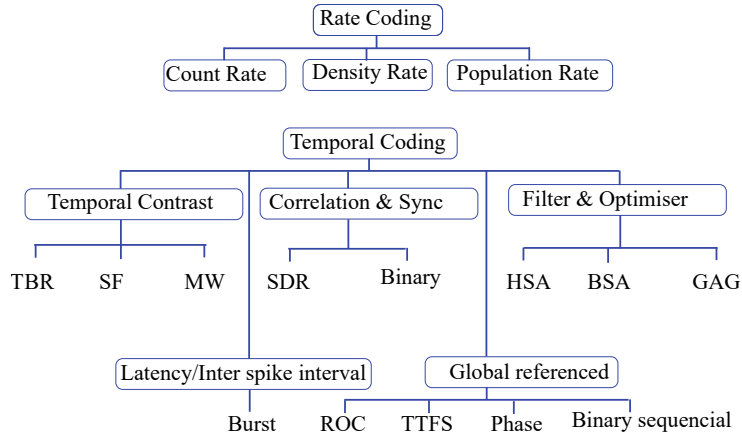}
\caption{Classification of rate and temporal coding methods}
\label{fig:stream}
\end{figure}
These coding schemes are typically applied to encode input data, such as visual information from a video camera, into spikes. A camera sensor, for example, might encode the intensity of light at each pixel as a series of spikes. In TTFS coding, a highly exposed pixel (one that receives a lot of light) would result in a quick spike, whereas in rate coding, more intense light would lead to a greater number of spikes within a given time frame. 

Building upon the distinctive gain-layer-on-waveguide technology, we propose a novel and efficient data encoding methodology tailored for photonic Spiking Neural Networks (SNNs). This encoding method capitalizes on threshold-based activation analogous to that observed in biological neurons, ensuring that energy consumption is minimized when the input remains below a defined threshold~\cite{zhang2024enhanced}. By exhibiting a threshold-driven behaviour similar to that employed in Time-To-First-Spike (TTFS) encoding, our approach leverages the precise timing of threshold crossing as a key informational variable. In other words, not only does the exact moment at which the input surpasses the threshold carry meaningful data—akin to temporal coding schemes—but, critically, the behaviour of the neuron-like element after firing also provides valuable insights into the signal’s strength and rate characteristics.

Unlike purely time-based encoding strategies, our method integrates both temporal and rate-based elements. The initial threshold crossing event conveys essential temporal information, while the subsequent modulation of firing rate supplies additional layers of encoded data. By merging these temporal and rate-based components, our encoding expands the representational capacity of photonic SNNs, enabling more nuanced and robust signal processing.

To demonstrate the practical advantages of this dual-mode encoding scheme, we introduce a photonic SNN architecture that implements the proposed encoding strategy for classification tasks. In this configuration, input signals are translated into spike-based representations governed by threshold-dependent, energy-efficient firing events. Subsequent neural layers interpret and refine these representations to achieve accurate classification performance. This combined approach—integrating advanced photonic hardware, threshold-based encoding, and biologically inspired spiking dynamics—provides a promising pathway towards high-speed, energy-efficient, and scalable photonic neuromorphic computing systems.

\subsection*{Threshold-triggered intensity-based encoding method}
In the proposed Threshold-triggered intensity-based encoding method, no spiking activity occurs as long as the integrated input stimulus remains below a biologically-inspired threshold level. This stands in contrast to many conventional encoding strategies, where any input typically produces some form of response. Here, once the input intensity surpasses the prescribed threshold, the system transitions into an active regime. In this active regime, two key changes are observed: first, the latency between the onset of the input pulse train and the first spike decreases as the input intensity increases; second, the amplitude of the emitted spikes grows proportionally with the input strength.

This dual dependence—on both the occurrence of a threshold and the graded modulation of spike response—enables the encoding mechanism to more faithfully replicate a broader range of neuronal behaviours. Unlike time-to-first-spike (TTFS) methods, which depend solely on spike latency to encode input strength, the Threshold-triggered intensity-based encoding scheme simultaneously leverages the presence of a threshold event and the strength of the resultant responses. As a result, the Threshold-triggered intensity-based encoding framework inherently captures richer information embedded in the stimulus, thereby providing an expanded representational space that conventional methods may overlook.
\begin{figure}[ht]
\centering
\includegraphics[scale=0.5]{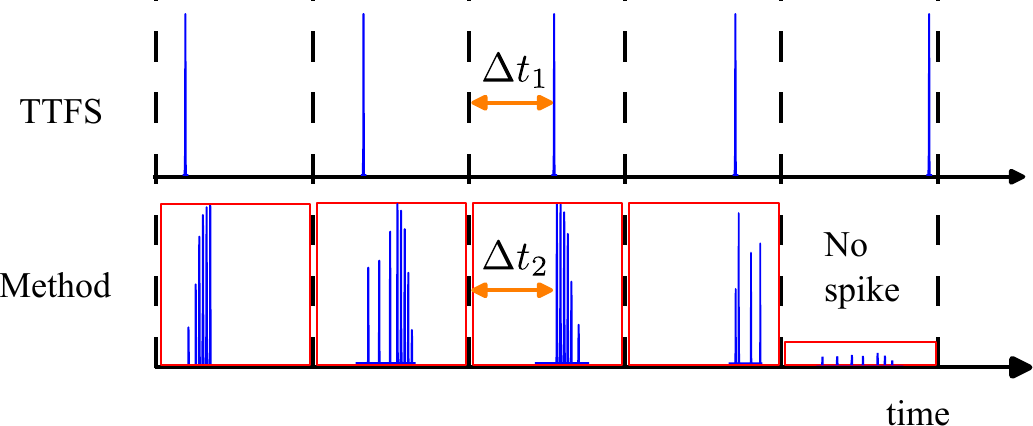}
\caption{Comparison of time-to-first-spike (TTFS) encoding with the proposed threshold-triggered intensity-based encoding method. Below a biologically-inspired threshold, no spikes occur, conserving energy. Once the threshold is surpassed, spike latency decreases and spike amplitude increases with input intensity. By jointly capturing threshold activation and graded response strength, this method provides richer information encoding and greater adaptability than conventional strategies.}
\label{fig:ComparisonbetweenTTFSandmethod}
\end{figure}
To illustrate the differences, a comparison between TTFS and this new Threshold-triggered intensity-based encoding approach is depicted in Fig.~\ref{fig:ComparisonbetweenTTFSandmethod}. When the input pulses vary in amplitude, frequency, or other parameters, the spike timing, the spike strength, and the very manifestation of threshold activation shift accordingly. This ensures that the neural encoding can adapt to a wide range of input conditions without unnecessary energy expenditure. Importantly, in the absence of input intensity exceeding the threshold, no spikes are generated and, therefore, no energy is consumed. This feature significantly enhances the energy efficiency of the Threshold-triggered intensity-based encoding method, making it a more practical and biologically plausible encoding strategy.

\section*{Discussion}

The development of the Doped-Gain-Layer-on-Waveguide-Synapse signifies a pivotal breakthrough in neuromorphic photonics, directly tackling the major hurdles faced by current spiking neural network (SNN) technologies. By integrating a rare-earth-doped gain layer onto a photonic waveguide, our approach effectively addresses critical issues such as optical losses, nonlinear computation challenges, and scalability limitations inherent in both opto-electronic and all-optical systems.

Our simulations, based on an Erbium-doped three-level model, reveal the dynamic capabilities of the Doped-Gain-Layer-on-Waveguide-Synapse essential for neuromorphic computing. The temporal behavior of ion populations across multiple energy levels demonstrates the system's proficiency in threshold-based spike generation, showcasing asynchronous spike initiation and significant signal amplification. Specifically, input pump and probe pulses modulate the doped layer's state potential, surpassing the transparency threshold and triggering nonlinear optical responses that closely emulate biological neuronal firing.

Figures \ref{fig:threshold_property__varyPeakpower} and \ref{fig:threshold_property__varypitch} illustrate the synapse's sensitivity to variations in input peak powers and inter-spike intervals, respectively. These results confirm the device's ability to exhibit threshold-dependent behavior and temporal integration—key for processing temporal data and encoding neural information. Additionally, the signal enhancement observed in Figure \ref{fig:overlay_input_and_output_sig_1} highlights the gain layer's role in amplifying input signals, thereby enhancing output spike fidelity and ensuring reliable signal transmission across the synapse.

The short-term memory capability, depicted in Figure \ref{fig:MERGED_SET1SET2tempmemoryproperty}, underscores the device's ability to retain state information over brief intervals, akin to synaptic plasticity in biological systems. This feature is crucial for tasks requiring temporal context and sequence recognition, significantly boosting the computational versatility of photonic neural networks.

The Doped-Gain-Layer-on-Waveguide-Synapse's capability to perform complex, nonlinear computations with enhanced energy efficiency and scalability positions it as a cornerstone for next-generation cognitive computing platforms. Leveraging the unparalleled speed and parallelism of photonic systems, our synapse enables the construction of highly interconnected neural networks capable of sophisticated cognitive tasks, including real-time data processing, pattern recognition, and adaptive learning.

Furthermore, the short-term memory and asynchronous operation capabilities imbue the synapse with functionalities reminiscent of biological neural networks, allowing for more nuanced and context-aware computations. This alignment with natural neuronal processes not only boosts computational efficiency but also paves the way for developing more intelligent and adaptive AI systems.

\subsection*{Conclusion}

In conclusion, the Doped-Gain-Layer-on-Waveguide-Synapse represents a groundbreaking innovation in neuromorphic photonics, providing a robust solution to the pressing challenges of computational power, energy efficiency, and scalability in neural network implementations. By integrating a rare-earth-doped gain layer onto a photonic waveguide, our approach as high proposes high-fidelity spike generation, larger SNN scaling, enhanced dynamic response capabilities, and efficient signal amplification within a compact on-chip footprint. The promising simulation results underscore the synapse's potential to emulate biological neuronal behaviors, laying the foundation for highly efficient and scalable photonic neural networks.

Looking ahead, the proposed synapse is set to significantly propel the field of neuromorphic computing forward, facilitating the realization of next-generation cognitive computing platforms that harness the speed and parallelism of photonic systems. Continued refinement of the design, optimization of performance, and resolution of existing limitations will ensure that the Doped-Gain-Layer-on-Waveguide-Synapse plays a pivotal role in the evolution of intelligent, adaptive, and energy-efficient AI technologies.

\section*{Acknowledgements}

The authors would like to thank the Eindhoven Hendrik Casimir Institute (EHCI) for their support. This work was supported by [\textcolor{red}{RF201750-10}].

\newpage
\bibliography{sample}

\end{document}